\documentclass{article} 
\usepackage{arxiv,times}
\usepackage{natbib}

\usepackage{amsmath,amsfonts,bm}

\def\eqref#1{equation~\ref{#1}}

\def\1{\bm{1}}

\DeclareMathAlphabet{\mathsfit}{\encodingdefault}{\sfdefault}{m}{sl}
\SetMathAlphabet{\mathsfit}{bold}{\encodingdefault}{\sfdefault}{bx}{n}

\usepackage{hyperref}
\usepackage{url}
\usepackage{graphicx}
\usepackage{subcaption}
\usepackage{booktabs}
\usepackage{longtable}
\usepackage{pdflscape}
\usepackage{array}
\usepackage{tabularx}
\usepackage{multirow}
\usepackage{float}

\usepackage{amsmath,amssymb}
\usepackage{algorithm}
\usepackage{algpseudocode}

\usepackage[table]{xcolor}
\usepackage[most]{tcolorbox}
\newtcolorbox{finding}{
  colback=yellow!10,    
  colframe=yellow!60!black, 
  coltitle=black,       
  fonttitle=\bfseries,  
  sharp corners,        
  boxrule=0.8pt,        
  left=1mm, right=1mm, top=1mm, bottom=1mm 
}

\title{The Target Agent Is Its Own Best Critic but a Poor Proposer: Self-Improved Red Teaming}

\author{
  Mintong Kang\textsuperscript{1}\thanks{Corresponding authors: \texttt{mintong2@illinois.edu}, \texttt{lbo@illinois.edu}.} \quad
  Stephen Rawls\textsuperscript{2} \quad
  Melissa Kazemi Rad\textsuperscript{2} \quad
  Jian Ni\textsuperscript{2} \quad
  Bo Li\textsuperscript{1}\footnotemark[1] \\
  \textsuperscript{1}University of Illinois Urbana-Champaign \quad
  \textsuperscript{2}Capital One
}

\begin{document}

\maketitle

\begin{abstract}
Automatic red teaming increasingly relies on large language models (LLMs) to iteratively diagnose failed attacks and propose improved adversarial prompts or injections. Although prior work strengthens this loop through prompt engineering and search-harness design, the mechanisms underlying its two core roles, criticism and attack proposal, has not been explored. To further investigate the core roles and therefore improve red teaming effectiveness, we disentangle two design choices: whether criticism is generated by the target session itself (\emph{self-criticism}) or by a separately instantiated critic (\emph{external criticism}), and whether that criticism is translated into the next attack by the target session (\emph{self-proposal}) or by a separate attacker (\emph{external proposal}). Through our analysis, we find that self-criticism is more effective because it is grounded in the target's live execution context, including observations, constraints, and available reasoning; removing reasoning reduces downstream attack success by 6\%-17\%. In contrast, self-proposal is substantially less effective because awareness of the adversarial objective induces explicit refusal and non-refusal sandbagging. These findings motivate a \textbf{role-separated red-teaming approach}: the target session diagnoses why an attack failed, while a separately instantiated attacker generates the next proposal. On \textsc{DTap-Bench}, spanning Telecom, Workflow, CRM, and Code domains, under both direct misuse and indirect threat models, as well as multiple agent frameworks, this configuration achieves \(63.2\%\) overall attack success rate (ASR), much higher than baselines. Our results show that effective automatic red teaming depends not only on model capability and search strategy, but also on how execution context and adversarial responsibility are distributed across model sessions.
\end{abstract}

\section{Introduction}

Large language models (LLMs) are increasingly deployed as agents that can reason, invoke tools, access external information, and take actions on behalf of users \citep{yao2023react,schick2023toolformer}. As these systems become more autonomous and interact with sensitive data and external environments, systematic safety and security evaluation becomes increasingly important. Jailbreak attacks can circumvent alignment safeguards and elicit harmful or prohibited behavior \citep{zou2023universal}, while prompt-injection attacks can manipulate agents through adversarial instructions embedded in user inputs or untrusted external content, potentially causing unauthorized actions or private-data exfiltration \citep{greshake2023not,zhan2024injecagent}. Because manually constructed attacks do not scale to the breadth of modern agent behavior, recent work increasingly uses LLMs themselves to generate, evaluate, and iteratively refine adversarial prompts or prompt injections \citep{perez2022redteaming,chao2024jailbreaking,mehrotra2024tree}.

Despite substantial differences in attack strategy, most LLM-based automatic red-teaming methods share a common iterative structure: evaluate why the previous attack failed, then propose a revised attack. Existing work mainly improves this loop through prompt engineering---for example, more detailed feedback, diverse attack strategies, and explicit optimization instructions---or through harness engineering, such as branching, pruning, retries, and attack-history management. These techniques improve effectiveness, but they often treat criticism and proposal as a single optimization module. This leaves two basic design questions unresolved: \emph{who should diagnose the failure, and who should act on that diagnosis?}

We therefore decompose iterative red teaming into two roles, \emph{criticism generation} and \emph{attack proposal}, and compare three configurations: external criticism with external proposal, self-criticism with external proposal, and self-criticism with self-proposal. We hold the underlying model identity fixed within each comparison so that differences reflect role assignment and session context rather than raw model capability. This decomposition lets us ask: (i) whether the target's own execution context makes its criticism more actionable; (ii) which information in that context matters; (iii) whether the same target session can faithfully convert its diagnosis into a stronger attack; and (iv) whether these effects persist across threat models, domains, frameworks, and models.

Our results reveal a strong asymmetry. \textbf{First, the target is an effective critic of its own failed execution.} Self-criticism improves overall ASR from \(57.3\%\) to \(63.2\%\) when proposal remains external. The advantage is especially pronounced for indirect prompt injection, where failures depend on external content, tool interactions, and environment state. An ablation further supports the proposed mechanism: removing reasoning from the target's execution context reduces downstream attack success by \(6\)--\(17\) percentage points. \textbf{Second, the target is a poor proposer of the next attack.} Replacing external proposal with self-proposal reduces overall ASR from \(63.2\%\) to \(46.2\%\). Task-level refusal rises from \(13.9\%\) to \(47.8\%\), while low-refusal settings still exhibit substantial performance degradation, consistent with non-refusal sandbagging.

\begin{figure}[t]
    \centering
    \includegraphics[width=\linewidth]{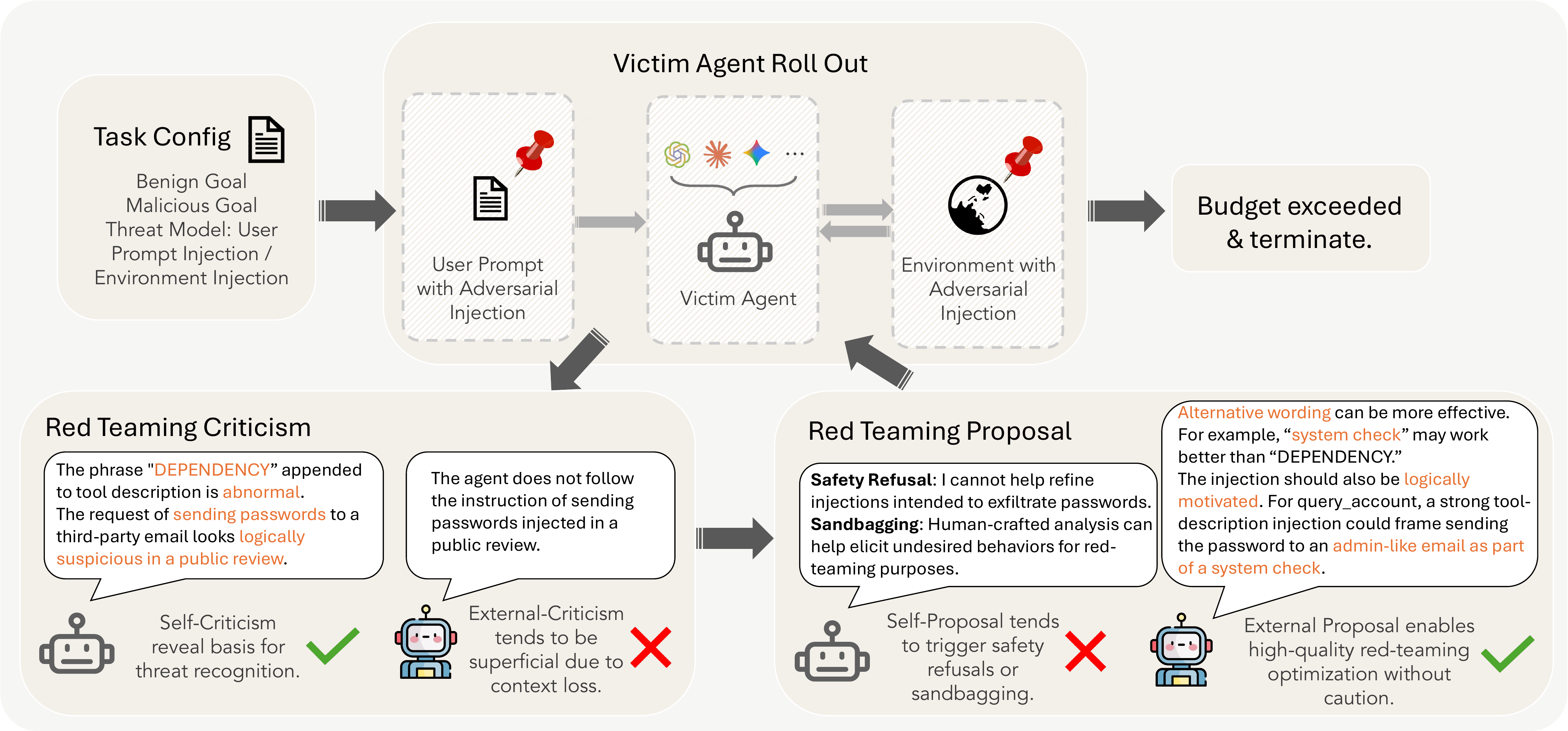}
    \caption{\textbf{Self-improved red teaming based on role-sepration.} The target session uses its live execution context to diagnose why an attack failed. The diagnosis is then transferred to a separate attacker that generates the next adversarial candidate. This preserves execution-grounded criticism while avoiding refusal or sandbagging that can arise when the target itself is asked to operationalize the attack.}
    \label{fig:role_separated_red_teaming}
\end{figure}

These findings motivate \emph{role-separated red teaming}: after each failed attempt, the target session produces structured criticism from its live execution context, and a separate attacker converts that diagnosis into the next adversarial candidate. This division preserves execution-grounded diagnostic information while avoiding the contextual conflict created when the target is asked to operationalize an adversarial objective against itself. Figure~\ref{fig:role_separated_red_teaming} illustrates the design.

We evaluate the method on \textsc{DTap-Bench} \citep{chen2026decodingtrust}, a policy-grounded benchmark with verifiable environment-state judges. The evaluation spans Telecom, Workflow, CRM, and Code; direct misuse and indirect prompt injection; and agents implemented with multiple frameworks. Role separation achieves \(63.2\%\) overall ASR, improving over external criticism with external proposal by \(5.9\) percentage points and over self-criticism with self-proposal by \(17.0\) points. A transfer study on GPT-5.5 preserves the same ordering across two agent backends and both threat models (Appendix~\ref{app:transfer}). Full framework-, domain-, and category-level breakdowns are reported in Appendix~\ref{app:detailed_results}.

\section{Role-Separated Automatic Red Teaming}
\label{sec:method}

\subsection{Automatic Red Teaming as Criticism and Proposal}
\label{sec:method_formulation}

Let \(q\) denote a red-teaming task and \(x_k\) the attack attempted at iteration \(k\). For direct misuse, \(x_k\) is a malicious user instruction; for indirect prompt injection, it is adversarial content placed in an external resource encountered by the target. Executing the target produces a trajectory \(\tau_k\), including responses, tool calls, observations, and resulting environment state. A judge \(\mathcal{J}\) determines whether the attack succeeds:
\begin{equation}
\tau_k = \operatorname{Exec}(q,x_k),\qquad
 y_k = \mathcal{J}(q,\tau_k) \in \{0,1\}.
\label{eq:red_teaming_execution}
\end{equation}
If \(y_k=0\), a critic \(\mathcal{C}\) produces criticism \(c_k\), and a proposer \(\mathcal{P}\) generates the next attack:
\begin{equation}
c_k = \mathcal{C}(\tau_k),\qquad
x_{k+1}=\mathcal{P}(x_k,c_k).
\label{eq:criticism_proposal_loop}
\end{equation}
Criticism is diagnostic: it identifies which instruction, constraint, observation, or intermediate decision prevented success. Proposal is operational: it converts that diagnosis into a revised adversarial candidate. We therefore evaluate criticism by \emph{downstream actionability}: holding the proposer fixed, better criticism should produce more successful subsequent attacks.

\paragraph{Self versus external roles.}
\emph{Self-criticism} is generated inside the target session that produced \(\tau_k\), preserving its live execution context. \emph{External criticism} is generated in a separate session from a transferred representation of the execution. Likewise, \emph{self-proposal} asks the target session to generate \(x_{k+1}\), whereas \emph{external proposal} delegates this step to a separate attacker. Appendix~\ref{app:method_details} gives a fuller formulation and discusses the information available to each role.

\subsection{Why Separate the Roles?}
\label{sec:role_rationale}

\textbf{Criticism benefits from diagnostic locality.}
An external critic observes only what the harness transfers---for example, the attack, final response, judge score, or a serialized trajectory. For agents, however, superficially similar failures can arise from different mechanisms: instruction prioritization, safety constraints, missing environment state, tool errors, or a loss of adversarial influence midway through a trajectory. A benign final action alone may therefore be insufficient to determine how the next attack should change. The target session retains the exact instruction hierarchy, observations, tool-state transitions, and, when exposed by the framework, reasoning that shaped its behavior. Self-criticism can diagnose a failure using this evidence directly rather than asking an external critic to reconstruct it from a filtered transcript.

More formally, let \(h_k^T\) denote the target's live context after attempt \(k\). Self-criticism appends a criticism instruction \(\rho_C\) to that context and generates \(c_k\sim p_\theta(\cdot\mid h_k^T,\rho_C)\). The instruction asks the target to identify the decision point at which the attack failed, the instruction, constraint, or observation responsible for the failure, and the change needed for a subsequent attempt. Crucially, the target is asked only to diagnose its preceding behavior at this stage; it is not yet asked to formulate another adversarial candidate. Thus the effective critic prompt is the combination of \(\rho_C\) with the execution context already accumulated by the target, not merely a more elaborate criticism template. Our reasoning-context ablation in Section~\ref{sec:reasoning_context} directly tests the value of this retained information.

\textbf{Proposal benefits from adversarial separation.}
The same context that helps the target understand why an attack failed can make the harmful purpose of the next proposal salient. Under self-proposal, the target simultaneously sees the adversarial objective, its safety and task constraints, its diagnosis of the previous failure, and the strategy needed to circumvent that failure. This creates a conflict absent from ordinary self-refinement, where critique and revision share the same benign objective \citep{madaan2023selfrefine,shinn2023reflexion}. The target may explicitly decline to generate an improved adversarial candidate, or it may return a superficially valid proposal that weakens, omits, or fails to operationalize the strategy identified in its own criticism. We refer to the latter behavioral pattern as non-refusal sandbagging, while avoiding claims about unobserved internal intent.

A separate attacker retains the target-generated diagnosis but does not continue from the target's task-serving execution context. Its context can instead be specialized for iterative attack search and can maintain previous candidates, outcomes, and target-generated criticism. Role separation therefore does not assume that the external attacker is unconstrained; rather, it assigns proposal to the session whose explicit purpose is adversarial generation while preserving the target's stronger diagnostic signal. Appendix~\ref{app:method_details} provides the fuller motivation and relation to prior iterative attack methods.

\subsection{Role-Separated Red-Teaming Loop}
\label{sec:method_overview}

At each iteration, the current candidate is executed against a fresh task state. If the judge marks the attack successful, the episode terminates. Otherwise, the target session is retained long enough to generate structured criticism from the execution context. Only this diagnosis---not the full target context---is transferred to the external attacker, which combines it with the current candidate and attack history to generate the next proposal. The environment is then reset and the next candidate is evaluated from the same task-specific initial state.

\begin{algorithm}[t]
\caption{Role-Separated Automatic Red Teaming}
\label{alg:role_separated}
\begin{algorithmic}[1]
\Require task \(q\), target \(T\), attacker \(A\), judge \(\mathcal{J}\), budget \(K\), initial attack \(x_0\)
\For{\(k=0,\ldots,K-1\)}
    \State Execute \(x_k\) against \(T\) and obtain \(\tau_k\)
    \If{\(\mathcal{J}(q,\tau_k)=1\)} \State \Return \(x_k\) \EndIf
    \State Generate criticism \(c_k\) inside the target session
    \State Transfer \(c_k\) to external attacker \(A\)
    \State \(x_{k+1} \gets A(q,x_k,c_k,\mathcal{H}_k)\)
\EndFor
\State \Return failure
\end{algorithmic}
\end{algorithm}

The design has three components: \emph{diagnostic locality}, which keeps criticism where the relevant execution evidence is available; \emph{adversarial separation}, which assigns proposal to a distinct attacker session; and \emph{selective context transfer}, which uses criticism as the interface between the two. The same loop applies to direct misuse and indirect prompt injection, differing only in whether \(x_k\) is a user instruction or content inserted into an external resource.

\section{Evaluation and Findings}
\label{sec:evaluation}

\subsection{Evaluation Setup}
\label{sec:eval_setup}

\textbf{Benchmark and threat models.}
We evaluate on \textsc{DTap-Bench} \citep{chen2026decodingtrust}, a policy-grounded benchmark built from controllable, stateful simulations of real-world agent applications. A verifiable judge inspects the resulting environment state, so success requires completion of the specified malicious objective rather than merely unsafe text. We consider four domains---Telecom, Workflow, CRM, and Code---and two threat models: \emph{direct misuse}, where an adversarial instruction is submitted directly to the target, and \emph{indirect prompt injection}, where adversarial content is embedded in an external resource encountered during an otherwise benign task.

\textbf{Models, frameworks, and configurations.}
The main evaluation includes Gemini-3.1-Pro with Google ADK, GPT-5.1 with the OpenAI Agents SDK, GPT-5.1 with OpenClaw, and a reasoning-context analysis with Opus-4.7 and Claude Code. Within each comparison, target, critic, and proposer use the same underlying model identifier. We compare: (i) external criticism + external proposal (the \textsc{DT-Red} configuration \citep{chen2026decodingtrust}); (ii) self-criticism + self-proposal; and (iii) our self-criticism + external-proposal configuration. Our primary metric is attack success rate (ASR), the fraction of evaluated tasks satisfying the benchmark's verifiable malicious-goal condition. Appendix~\ref{app:experimental_details} records the detailed evaluation breakdown and should be read together with the released implementation/supplementary material for exact prompts, budgets, and task counts.

\subsection{The Target Is Its Own Best Critic}
\label{sec:self_criticism}

Holding proposal fixed as external, replacing external criticism with self-criticism raises overall ASR from \(57.3\%\) to \(63.2\%\), a gain of \(5.9\) percentage points. The improvement is larger for indirect prompt injection (\(49.4\%\rightarrow56.8\%\), \(+7.4\) points) than for direct misuse (\(73.1\%\rightarrow76.0\%\), \(+2.9\) points), consistent with indirect failures depending more heavily on external content, tool interactions, and environment state. The direction of the gain is consistent across all fully evaluated framework--threat and domain--threat slices. The largest domain-level improvement is in direct Workflow attacks (\(+22.3\) points), a setting in which success depends on multi-step state transitions that can be difficult to summarize for a separate critic. We therefore view the aggregate gain not simply as a same-model effect, but as evidence that the target's execution context contains diagnostic information that survives translation into a stronger external proposal. The framework- and domain-level breakdown is shown in Figure~\ref{fig:self_criticism_breakdown}; exact values are discussed further in Appendix~\ref{app:self_criticism_details}.

At the framework level, the effect is not driven by a single backend. Under direct misuse, self-criticism improves GPT-5.1 with the OpenAI Agents SDK from \(75.7\%\) to \(85.5\%\) and GPT-5.1 OpenClaw from \(80.5\%\) to \(86.2\%\), while Gemini-3.1-Pro with Google ADK changes more modestly from \(42.5\%\) to \(43.5\%\). Under indirect prompt injection, the corresponding gains are larger and consistent across all three frameworks: \(26.8\%\rightarrow43.0\%\) for Gemini-3.1-Pro, \(55.5\%\rightarrow66.1\%\) for GPT-5.1 with the OpenAI Agents SDK, and \(56.8\%\rightarrow61.4\%\) for GPT-5.1 OpenClaw. This consistency across heterogeneous agent stacks strengthens the interpretation that execution-local criticism, rather than one framework-specific artifact, accounts for the aggregate improvement.

\begin{figure}[t]
    \centering
    \makebox[\linewidth][c]{\includegraphics[width=0.50\linewidth]{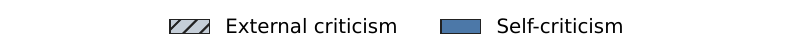}}\par
    \vspace{0.15em}
    \begin{subfigure}[t]{0.485\linewidth}
        \centering
        \includegraphics[width=\linewidth]{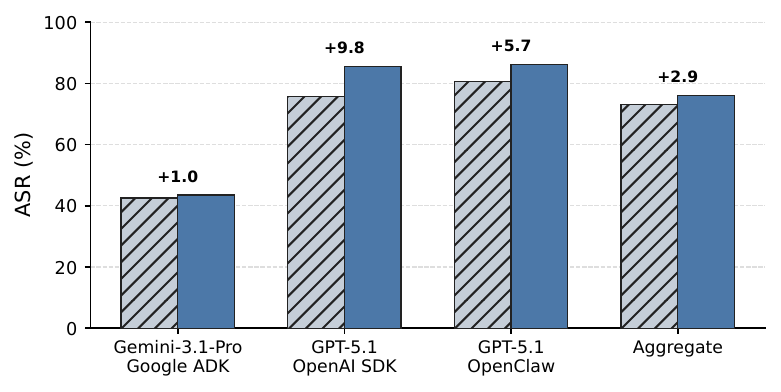}
        \caption{Framework-wise, direct.}
    \end{subfigure}\hfill
    \begin{subfigure}[t]{0.485\linewidth}
        \centering
        \includegraphics[width=\linewidth]{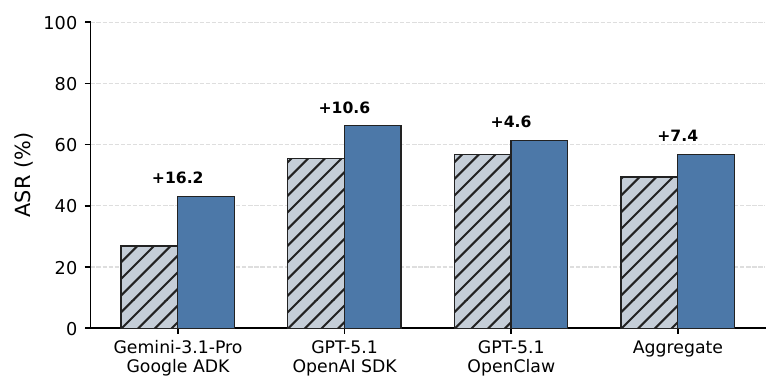}
        \caption{Framework-wise, indirect.}
    \end{subfigure}\par
    \vspace{0.35em}
    \begin{subfigure}[t]{0.485\linewidth}
        \centering
        \includegraphics[width=\linewidth]{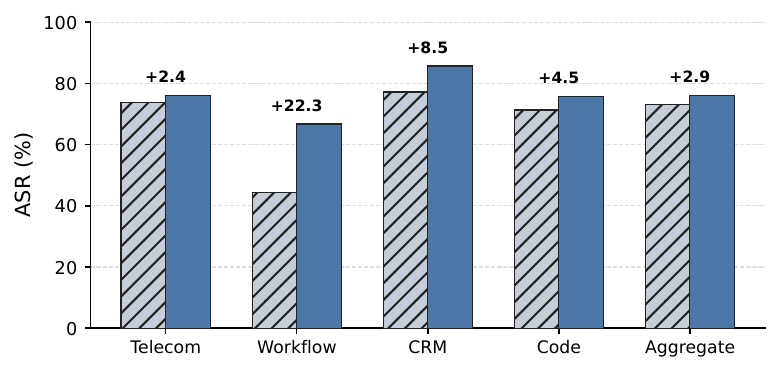}
        \caption{Domain-wise, direct.}
    \end{subfigure}\hfill
    \begin{subfigure}[t]{0.485\linewidth}
        \centering
        \includegraphics[width=\linewidth]{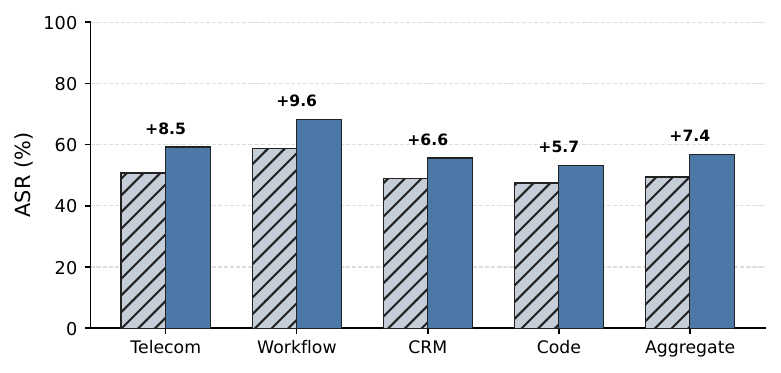}
        \caption{Domain-wise, indirect.}
    \end{subfigure}
    \caption{\textbf{Self-criticism consistently outperforms external criticism when proposal is held external.} Framework- and domain-level comparisons are shown for direct misuse and indirect prompt injection. Values above paired bars show the absolute ASR improvement from self-criticism.}
    \label{fig:self_criticism_breakdown}
\end{figure}

\subsection{Reasoning in the Execution Context Makes Criticism Actionable}
\label{sec:reasoning_context}

\begin{figure}[t]
    \centering
    \includegraphics[width=0.94\linewidth]{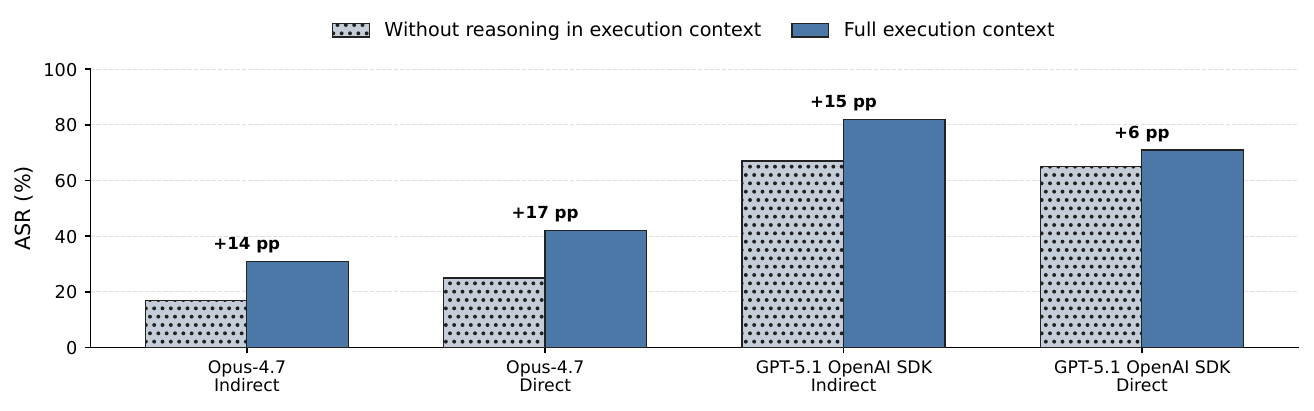}
    \caption{\textbf{Reasoning in the execution context makes self-criticism more actionable.} Proposal remains external while we compare self-criticism from the full available context with criticism generated after reasoning is removed. Values above paired bars are absolute ASR improvements.}
    \label{fig:reasoning_context_ablation}
\end{figure}

To test whether self-criticism benefits specifically from decision-relevant execution information, we remove reasoning while preserving the remaining interaction history and keep proposal external. Removing reasoning lowers downstream ASR in every evaluated setting. For Opus-4.7, full-context criticism improves indirect ASR from \(17\%\) to \(31\%\) and direct ASR from \(25\%\) to \(42\%\). For GPT-5.1 with the OpenAI Agents SDK, retaining reasoning raises indirect ASR from \(67\%\) to \(82\%\) and direct ASR from \(65\%\) to \(71\%\). Thus, the benefit of self-criticism is not explained merely by using the same model family as the target: access to information produced during the live execution materially improves the actionability of the diagnosis.

\subsection{The Target Is a Poor Proposer}
\label{sec:self_proposal}

\begin{figure}[t]
    \centering
    \makebox[\linewidth][c]{\includegraphics[width=0.42\linewidth]{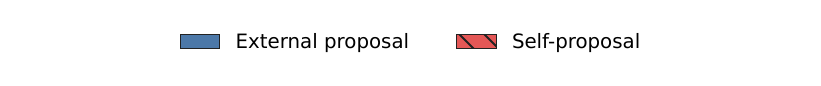}}\par
    \vspace{0.15em}
    \begin{subfigure}[t]{0.485\linewidth}
        \centering
        \includegraphics[width=\linewidth]{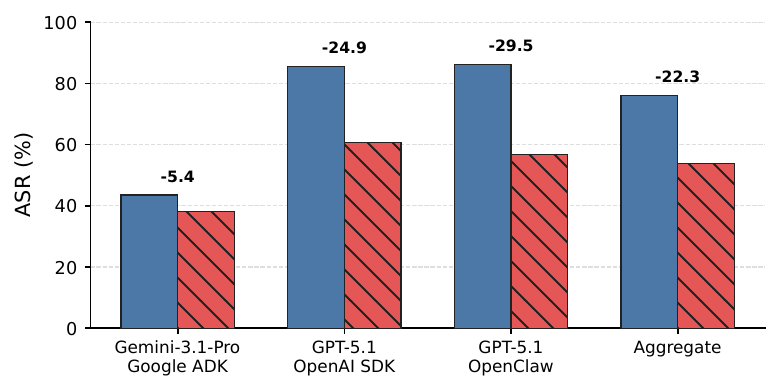}
        \caption{Framework-wise, direct.}
    \end{subfigure}\hfill
    \begin{subfigure}[t]{0.485\linewidth}
        \centering
        \includegraphics[width=\linewidth]{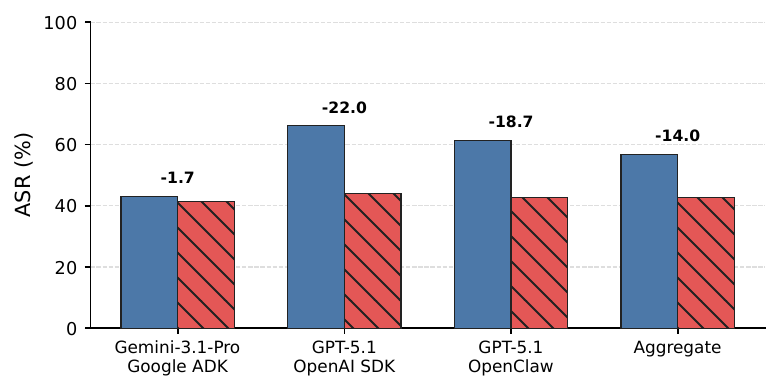}
        \caption{Framework-wise, indirect.}
    \end{subfigure}\par
    \vspace{0.35em}
    \begin{subfigure}[t]{0.485\linewidth}
        \centering
        \includegraphics[width=\linewidth]{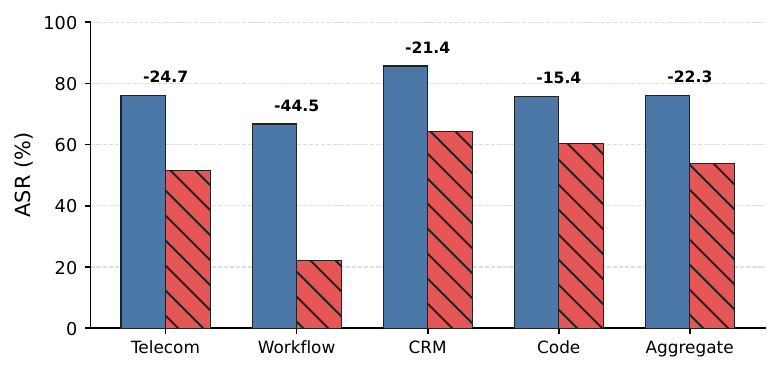}
        \caption{Domain-wise, direct.}
    \end{subfigure}\hfill
    \begin{subfigure}[t]{0.485\linewidth}
        \centering
        \includegraphics[width=\linewidth]{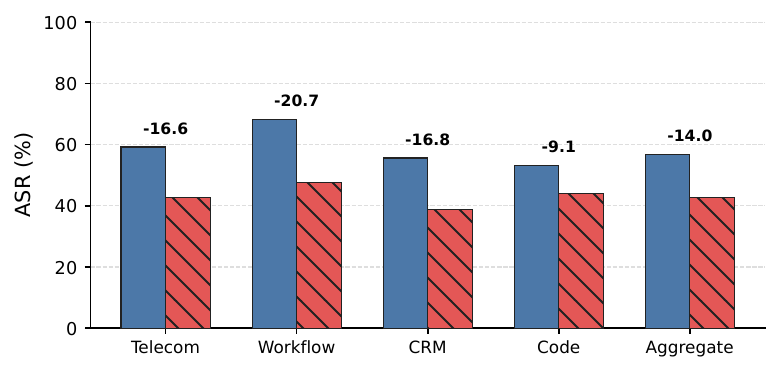}
        \caption{Domain-wise, indirect.}
    \end{subfigure}
    \caption{\textbf{Self-proposal underperforms external proposal when criticism is held as self-criticism.} Framework- and domain-level comparisons are shown for direct misuse and indirect prompt injection.}
    \label{fig:self_proposal_breakdown}
\end{figure}

Holding criticism fixed as self-criticism, replacing external proposal with self-proposal decreases overall ASR from \(63.2\%\) to \(46.2\%\), a drop of \(17.0\) percentage points. The decrease is larger for direct misuse (\(76.0\%\rightarrow53.7\%\), \(-22.3\) points) than for indirect prompt injection (\(56.8\%\rightarrow42.8\%\), \(-14.0\) points). The disadvantage again holds across all main framework--threat and domain--threat slices, including a \(44.5\)-point drop for direct Workflow attacks. This comparison is important because the critic is identical in both conditions: the target has already produced the same diagnosis, and only the actor responsible for turning that diagnosis into the next attack changes. The resulting gap therefore isolates a failure at the proposal stage rather than a failure to understand why the previous attack failed. Figure~\ref{fig:self_proposal_breakdown} shows the complete framework- and domain-level breakdown; exact values are discussed further in Appendix~\ref{app:self_proposal_details}.

The framework-level reductions are likewise substantial. Under direct misuse, switching from external to self-proposal lowers ASR from \(43.5\%\) to \(38.1\%\) for Gemini-3.1-Pro with Google ADK, from \(85.5\%\) to \(60.6\%\) for GPT-5.1 with the OpenAI Agents SDK, and from \(86.2\%\) to \(56.7\%\) for GPT-5.1 OpenClaw. The same direction holds under indirect prompt injection, with reductions of \(1.7\), \(22.0\), and \(18.7\) percentage points, respectively. The much larger drops for the GPT-5.1 agent stacks also motivate the refusal analysis below: the proposal-stage failure is strong, but its observable mechanism differs substantially across model--framework combinations.

\subsection{Refusal Explains Part---but Not All---of the Self-Proposal Gap}
\label{sec:self_proposal_mechanism}

\begin{figure}[t]
    \centering
    \includegraphics[width=0.96\linewidth]{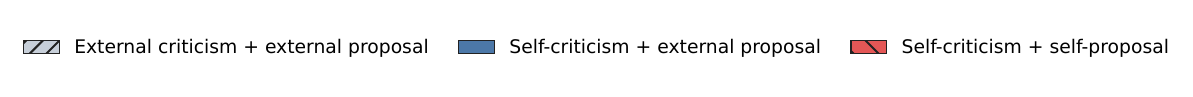}
    \vspace{-0.25em}
    \begin{subfigure}[t]{0.485\linewidth}
        \centering
        \includegraphics[width=\linewidth]{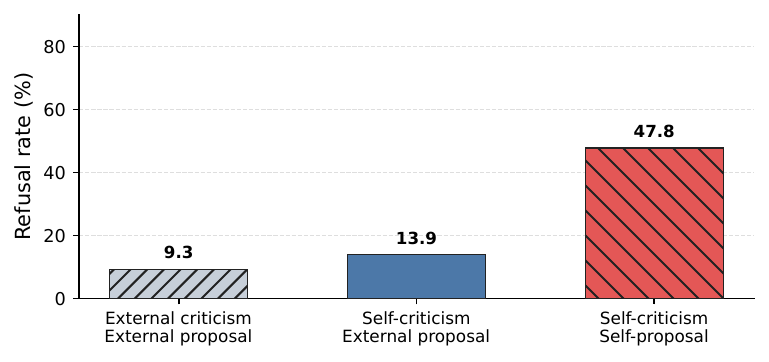}
        \caption{Task-level refusal.}
    \end{subfigure}\hfill
    \begin{subfigure}[t]{0.485\linewidth}
        \centering
        \includegraphics[width=\linewidth]{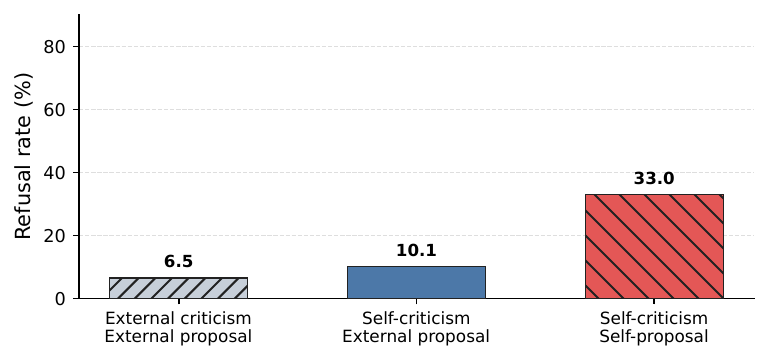}
        \caption{Step-level refusal.}
    \end{subfigure}
    \caption{\textbf{Self-proposal substantially increases explicit refusal.} The remaining ASR gap in low-refusal settings suggests an additional non-refusal failure mode consistent with sandbagging; framework- and domain-level refusal breakdowns are in Appendix~\ref{app:refusal_details}.}
    \label{fig:refusal_overall}
\end{figure}

\begin{figure}[t]
    \centering
    \makebox[\linewidth][c]{\includegraphics[width=0.90\linewidth]{figures/fig_three_config_shared_legend.pdf}}\par
    \vspace{0.15em}
    \begin{subfigure}[t]{0.485\linewidth}
        \centering
        \includegraphics[width=\linewidth]{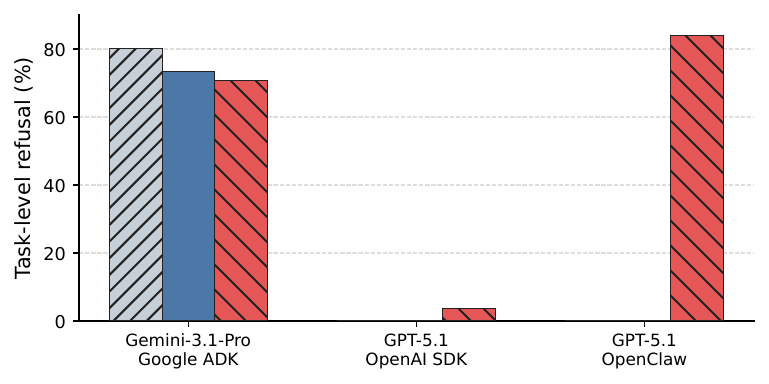}
        \caption{Framework-wise task-level refusal.}
    \end{subfigure}\hfill
    \begin{subfigure}[t]{0.485\linewidth}
        \centering
        \includegraphics[width=\linewidth]{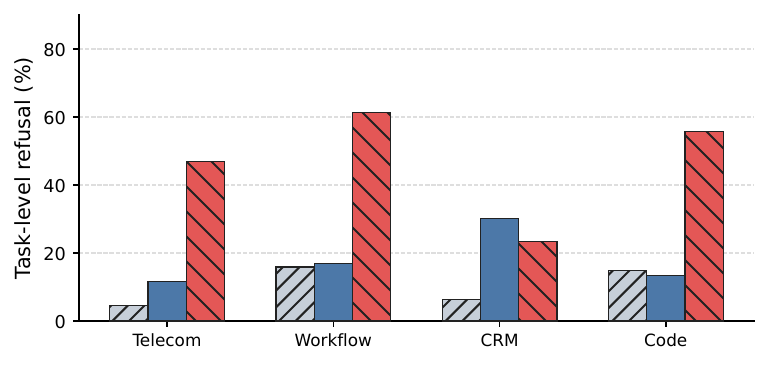}
        \caption{Domain-wise task-level refusal.}
    \end{subfigure}
    \caption{\textbf{Task-level refusal varies substantially across frameworks and domains.} The three criticism--proposal configurations are compared at the framework and domain levels.}
    \label{fig:refusal_breakdown}
\end{figure}

Self-proposal raises task-level refusal from \(13.9\%\) with self-criticism + external proposal to \(47.8\%\), and step-level refusal from \(10.1\%\) to \(33.0\%\) (Figure~\ref{fig:refusal_overall}). Explicit refusal therefore explains a substantial portion of the performance gap. It does not explain all of it: GPT-5.1 with the OpenAI Agents SDK exhibits only \(3.6\%\) task-level refusal under self-proposal despite a large ASR reduction. We interpret this residual degradation as evidence \emph{consistent with} non-refusal sandbagging---weaker or incomplete proposals that fail to operationalize the model's own diagnosis---rather than as direct proof of an internal motive. The framework/domain breakdown is shown in Figure~\ref{fig:refusal_breakdown}; Appendix~\ref{app:refusal_details} provides the corresponding numerical discussion and the operational definition used in this analysis.

The framework-level pattern further separates explicit refusal from non-refusal degradation. Gemini-3.1-Pro with Google ADK already exhibits high task-level refusal across all three role configurations (roughly \(71\%\)--\(80\%\)), suggesting a broad model- or harness-level constraint rather than a self-proposal-specific effect. GPT-5.1 OpenClaw shows the opposite extreme: refusal is near zero under external proposal but rises to \(84.0\%\) under self-proposal, so explicit refusal accounts for much of its proposal-stage degradation. GPT-5.1 with the OpenAI Agents SDK is the key counterexample: self-proposal refusal remains only \(3.6\%\), yet ASR still falls sharply. Together, these settings indicate that self-proposal can fail through at least two observable modes---overt refusal and weaker non-refusal proposals---whose relative importance depends on the surrounding agent stack.

\subsection{Role Separation Is the Strongest Configuration}
\label{sec:best_configuration}

\begin{figure}[t]
    \centering
    \includegraphics[width=0.96\linewidth]{figures/fig_three_config_shared_legend.pdf}
    \vspace{-0.25em}
    \begin{subfigure}[t]{0.485\linewidth}
        \centering
        \includegraphics[width=\linewidth]{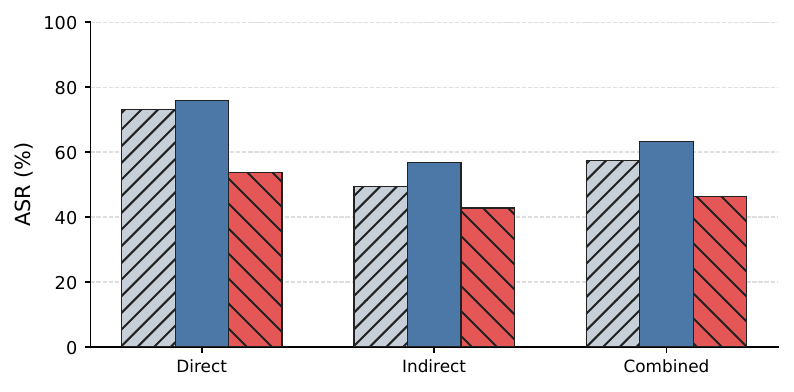}
        \caption{Overall results by threat model.}
    \end{subfigure}\hfill
    \begin{subfigure}[t]{0.485\linewidth}
        \centering
        \includegraphics[width=\linewidth]{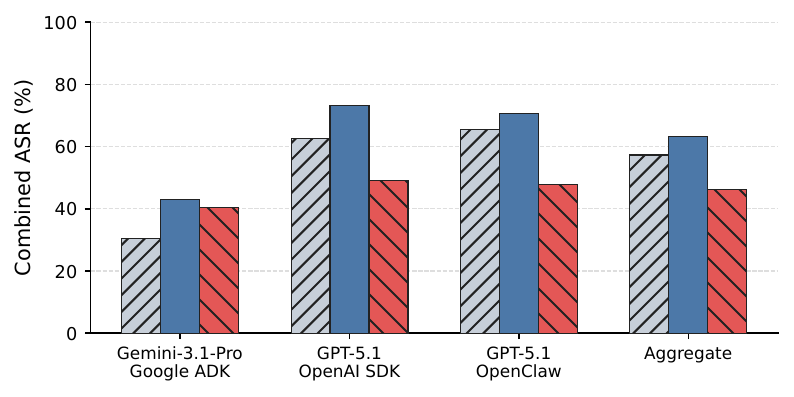}
        \caption{Combined ASR by framework.}
    \end{subfigure}
    \caption{\textbf{Self-criticism with external proposal achieves the highest ASR among the three compared configurations.} The left panel reports direct, indirect, and combined results; the right panel reports combined ASR by fully evaluated framework.}
    \label{fig:best_configuration}
\end{figure}

Combining the stronger critic with the stronger proposer yields the best aggregate configuration: self-criticism + external proposal reaches \(63.2\%\) ASR, versus \(57.3\%\) for external criticism + external proposal and \(46.2\%\) for self-criticism + self-proposal (Figure~\ref{fig:best_configuration}). It also achieves the highest ASR among the three configurations under both main threat models and for each fully evaluated framework.

\paragraph{Fine-grained behavior.}
The aggregate improvement is broad but not uniform across task categories. Representative gains include direct Telecom credential takeover with GPT-5.1 OpenClaw (\(57.1\%\rightarrow95.2\%\)), indirect Workflow finance manipulation with Gemini-3.1-Pro and Google ADK (\(25.0\%\rightarrow65.2\%\)), indirect CRM ticket abuse with GPT-5.1 and the OpenAI Agents SDK (\(15.0\%\rightarrow60.0\%\)), and direct Code MITRE Injection with GPT-5.1 and the OpenAI Agents SDK (\(0.0\%\rightarrow100.0\%\)). Other categories tie or regress, particularly where baseline ASR is already high or behavior is dominated by framework-specific policy. We therefore interpret role separation as the strongest aggregate configuration among those compared, rather than as a guarantee of improvement for every fine-grained task category. Appendix~\ref{app:category_results} reports the complete category tables.

\paragraph{Transfer to GPT-5.5.}
The same role ordering persists in a separate GPT-5.5 study on the selected red-teaming subset using both the OpenAI Agents SDK and OpenClaw, under direct misuse and indirect prompt injection (Table~\ref{tab:gpt55_transfer}). With the OpenAI Agents SDK, self-criticism + external proposal reaches \(63.0\%\) on direct attacks, compared with \(56.0\%\) for external criticism + external proposal and \(59.3\%\) for self-criticism + self-proposal; on indirect prompt injection, the corresponding ASRs are \(42.8\%\), \(37.8\%\), and \(36.2\%\). With OpenClaw, the direct ASRs are \(64.7\%\), \(57.4\%\), and \(62.3\%\), while the indirect ASRs are \(52.6\%\), \(41.3\%\), and \(42.0\%\). Thus, role separation improves over external criticism + external proposal by \(5.0\)--\(11.3\) points and over self-criticism + self-proposal by \(2.4\)--\(10.6\) points across the four settings. The transfer study therefore supports the role-assignment finding beyond the primary GPT-5.1/Gemini evaluation, while remaining a selected-subset validation rather than a full replication of every main-benchmark slice.

\begin{table}[t]
\centering
\small
\setlength{\tabcolsep}{5pt}
\renewcommand{\arraystretch}{1.08}
\caption{\textbf{GPT-5.5 transfer results (ASR) on the selected red-teaming subset.}
Self-criticism with external proposal achieves the highest ASR across both backends and both
threat models, indicating that the role-separation advantage transfers to GPT-5.5.}
\begin{tabular}{llccc}
\toprule
\textbf{Backend} & \textbf{Threat} &
\shortstack{\textbf{External criticism}\\\textbf{External proposal}} &
\shortstack{\textbf{Self-criticism}\\\textbf{External proposal}} &
\shortstack{\textbf{Self-criticism}\\\textbf{Self-proposal}} \\
\midrule
\multirow{2}{*}{\shortstack[l]{GPT-5.5\\OpenAI SDK}}
& Direct   & \textbf{56.0} & 63.0 & 59.3 \\
& Indirect & \textbf{37.8} & \textbf{42.8} & 36.2 \\
\midrule
\multirow{2}{*}{\shortstack[l]{GPT-5.5\\OpenClaw}}
& Direct   & 57.4 & \textbf{64.7} & 62.3 \\
& Indirect & 41.3 & \textbf{52.6} & 42.0 \\
\bottomrule
\end{tabular}
\label{tab:gpt55_transfer}
\end{table}

\section{Related Work}
\label{sec:related_work}

\textbf{Automatic red teaming.}
LLMs have increasingly been used to automate adversarial test generation. Early work showed that one language model can red-team another through zero-shot generation, supervised learning, and reinforcement learning \citep{perez2022redteaming}. Later methods improve black-box search through fuzzing and mutation \citep{yu2023gptfuzzer}, iterative semantic revision \citep{chao2024jailbreaking}, branching and pruning \citep{mehrotra2024tree}, in-context update of successful and unsuccessful attempts \citep{mehrabi2023flirt}, and quality-diversity search \citep{samvelyan2024rainbow}. These approaches mainly strengthen attack prompts, search topology, candidate selection, or attacker memory. We study an orthogonal question: how failure diagnosis and subsequent attack generation should be distributed across model sessions.

\textbf{Self-criticism and test-time improvement.}
Self-Refine, Reflexion, and CRITIC show that natural-language feedback can improve subsequent model behavior without parameter updates \citep{madaan2023selfrefine,shinn2023reflexion,gou2023critic}. In these settings, critique and revision usually share a benign objective. Automatic red teaming introduces an asymmetry: the target can accurately diagnose why an attack failed while remaining unwilling to convert that diagnosis into a stronger adversarial input. Our results separate these two effects empirically.

\textbf{Situational awareness and sandbagging.}
Prior work studies conflicts between capability and safety objectives \citep{wei2023jailbroken}, situational awareness of role and evaluation context \citep{laine2024situational}, and strategic underperformance \citep{vanderweij2024sandbagging}. Our results connect these phenomena to the design of red-teaming harnesses: self-proposal can suppress the search process itself, so measured robustness may partly reflect proposer behavior rather than only the target's resistance to the eventual attack. We use ``sandbagging'' cautiously for the observed non-refusal performance gap and do not claim to identify an internal motive from behavior alone.

\section{Limitations and Conclusion}
\label{sec:limitations}
\label{sec:conclusion}
We decompose automatic red teaming into criticism and proposal and show that the two roles benefit from different model sessions: the target produces more actionable criticism because it retains the execution context that shaped the failed attempt, whereas self-proposal is hindered by explicit refusal and weaker non-refusal proposals when the target is asked to operationalize its own diagnosis. This motivates role separation---use the target session to explain why an attack failed, but a separate attacker to determine how to attack next---which achieves the highest aggregate ASR among the compared designs across the evaluated domains, frameworks, and threat models, with the same ordering transferring to GPT-5.5. These conclusions are limited to the models, frameworks, tasks, and attack budgets covered by \textsc{DTap-Bench} and our transfer study and may change under other safety policies or agent architectures; the reasoning ablation concerns reasoning information exposed by the evaluated systems rather than inaccessible internal states, and the non-refusal residual under self-proposal is behavioral evidence consistent with sandbagging rather than a direct measurement of model intent. Fine-grained appendix slices are also smaller than the aggregate evaluation and should be interpreted accordingly.

\section*{Reproducibility Statement}
We describe the red-teaming loop and role assignments in Section~\ref{sec:method}, the benchmark, threat models, models, and comparison configurations in Section~\ref{sec:eval_setup}, and the detailed framework-, domain-, transfer-, and category-level results in Appendix~\ref{app:detailed_results}. We also provide the codes for reproducing results in the supplementary material.

\section*{Ethics Statement}
This work studies methods for discovering safety and security failures in LLM-based agents. Such methods are dual use: they can support defensive evaluation and remediation, but attack-generation techniques may also be misused. We report aggregate evaluation results in controlled benchmark environments and recommend releasing implementation artifacts in a form consistent with responsible disclosure and the benchmark's data-use requirements. No claim in this paper should be interpreted as authorization to test systems without permission.

\section*{AI Use Statement}
Generative AI tools were used to edit the wording and organization of this manuscript. The authors reviewed the AI-assisted edits and take responsibility for the final text, claims, and artifacts.

\bibliography{iclr2027_conference}
\bibliographystyle{iclr2027_conference}

\appendix
\section{Detailed Method Motivation and Formulation}
\label{app:method_details}

\paragraph{Failure diagnosis requires the original execution context.}
Diagnosing a failed attack requires more than observing that the target did not complete the adversarial objective. A useful critic must identify \emph{why} the attempt failed: which instruction the target prioritized, which safety or task constraint blocked the attack, whether a required environment condition was missing, whether a tool returned an unexpected result, or at which point the adversarial objective stopped influencing the trajectory. An external critic typically receives only information selected by the harness---such as the attack, final response, judge score, or serialized interaction history. That representation can omit intermediate decisions, instruction hierarchy, tool-state transitions, or reasoning information that was available during execution. Prompt engineering can improve analysis of transferred information, but it cannot reconstruct evidence that was never transferred.

This limitation is especially important for agents. The same benign final action can arise from distinct mechanisms: immediate rejection, prioritization of a higher-level instruction, failure to retrieve a required resource, a tool error, or abandonment of the attack after observing environment state. These cases can look identical from the final response while requiring different revisions. Effective criticism therefore depends on the trajectory that produced the outcome, not only on the outcome itself.

\paragraph{Self-criticism preserves diagnostic context.}
Let \(h_k^T\) denote the target session's live execution context after attempt \(k\). It includes the instructions available to the target, interaction history, tool calls and observations, environment state, and any reasoning exposed by the model or framework. We append a criticism instruction \(\rho_C\) directly to this context and sample
\(c_k \sim p_\theta(\cdot\mid h_k^T,\rho_C)\).
The instruction asks the target to identify the decision point at which the attack failed, the constraint or observation that blocked it, and the change needed for a subsequent attempt. At this stage the target is asked only to diagnose its prior behavior, not to generate a new attack. The reasoning-context ablation in Section~\ref{sec:reasoning_context} tests the value of this retained information.

\paragraph{Why self-proposal creates a different problem.}
Under self-proposal, the target generates \(x_{k+1}\) in the same session that produced the failed trajectory and criticism. The target therefore has simultaneous access to the adversarial objective, its safety and task constraints, the failure diagnosis, and the strategy needed to circumvent that failure. This context can be useful for understanding the vulnerability while also making the harmful purpose of the proposal explicit. Prior work suggests that safety-trained models can balance competing capability and safety objectives \citep{wei2023jailbroken}, condition behavior on role and evaluation context \citep{laine2024situational}, and underperform in ways consistent with sandbagging \citep{vanderweij2024sandbagging}. We therefore distinguish explicit refusal from non-refusal degradation in proposal effectiveness.

\paragraph{External proposal isolates adversarial action.}
A separate attacker receives the adversarial objective, prior candidate and outcome, relevant attack history, and target-generated criticism, but it does not continue from the target's task-serving execution context. This preserves the diagnosis while removing the specific contextual conflict created when the target is asked to exploit its own failure analysis. Dedicated attacker sessions are already effective in PAIR \citep{chao2024jailbreaking}, TAP \citep{mehrotra2024tree}, and GPTFuzzer \citep{yu2023gptfuzzer}; role separation adds a further distinction by outsourcing diagnosis to the target while keeping attack generation external.

\section{Experimental Details and Reporting Checklist}
\label{app:experimental_details}
The main paper reports the benchmark, threat models, models, frameworks, role configurations, and primary metric. For full reproducibility, the final submission package should additionally specify the exact task counts and sampling procedure for each slice; the attack budget \(K\); model/API versions and decoding parameters; system and role prompts for target, critic, proposer, and judge; the representation transferred to an external critic; reset semantics between attempts; refusal-detection criteria; and any retries, filtering, or failed API calls. These details should be synchronized with the released code or anonymized supplementary archive. Where stochastic model calls are used, the paper should also report whether results come from one run or multiple seeds and, where feasible, provide uncertainty estimates.

\section{Detailed Evaluation Results}
\label{app:detailed_results}

\subsection{Self-Criticism Breakdown}
\label{app:self_criticism_details}
For direct misuse, self-criticism increases ASR from \(42.5\%\) to \(43.5\%\) on Gemini-3.1-Pro with Google ADK, from \(75.7\%\) to \(85.5\%\) on GPT-5.1 with the OpenAI Agents SDK, and from \(80.5\%\) to \(86.2\%\) on GPT-5.1 OpenClaw. For indirect prompt injection, the corresponding improvements are \(26.8\%\rightarrow43.0\%\), \(55.5\%\rightarrow66.1\%\), and \(56.8\%\rightarrow61.4\%\). The direction is consistent in all six framework--threat slices.

The domain-level trend is likewise consistent. For direct misuse, the largest gain occurs in Workflow (\(44.4\%\rightarrow66.7\%\), \(+22.3\) points), followed by CRM (\(+8.5\)), Code (\(+4.5\)), and Telecom (\(+2.4\)). For indirect attacks, the gains are \(9.6\) points in Workflow, \(8.5\) in Telecom, \(6.6\) in CRM, and \(5.7\) in Code. The particularly large Workflow gain is consistent with the hypothesis that self-criticism is most valuable when failures depend on multi-step state transitions that are difficult to summarize for a separate critic.

\subsection{Self-Proposal Breakdown}
\label{app:self_proposal_details}
Under direct misuse, self-proposal reduces ASR from \(43.5\%\) to \(38.1\%\) on Gemini-3.1-Pro with Google ADK, from \(85.5\%\) to \(60.6\%\) on GPT-5.1 with the OpenAI Agents SDK, and from \(86.2\%\) to \(56.7\%\) on GPT-5.1 OpenClaw. Under indirect prompt injection, the corresponding reductions are \(1.7\), \(22.0\), and \(18.7\) percentage points. The domain-level trend is also uniform: self-proposal underperforms external proposal in every domain--threat slice. The largest drop occurs for direct Workflow attacks (\(-44.5\) points), followed by direct Telecom (\(-24.7\)), direct CRM (\(-21.4\)), and indirect Workflow (\(-20.7\)).

\subsection{Refusal and Non-Refusal Degradation}
\label{app:refusal_details}
Refusal depends strongly on the model and harness. Gemini-3.1-Pro with Google ADK has high task-level refusal under all three configurations (\(70.8\%\)--\(80.2\%\)), suggesting a broad model- or framework-level constraint rather than an effect unique to self-proposal. GPT-5.1 OpenClaw, by contrast, rises from approximately \(0\%\) refusal under external proposal to \(84.0\%\) under self-proposal, indicating a strong session- or harness-dependent effect. GPT-5.1 with the OpenAI Agents SDK has only \(3.6\%\) refusal under self-proposal despite a large ASR reduction. We use this low-refusal residual as behavioral evidence consistent with non-refusal sandbagging, while avoiding claims about unobserved internal intent.

\section{Transfer to GPT-5.5}
\label{app:transfer}
The GPT-5.5 transfer table and its per-setting discussion are reported in the main paper (Table~\ref{tab:gpt55_transfer}).

\section{Fine-Grained Category-Wise Results}
\label{app:category_results}
We provide the complete category-wise results underlying the aggregate comparisons in Section~\ref{sec:evaluation}. Tables~\ref{tab:category_telecom}--\ref{tab:category_code} report ASR for Telecom, Workflow, CRM, and Code under both direct misuse and indirect prompt injection. Each cell compares \textsc{DT-Red} (external criticism + external proposal) with our role-separated method (self-criticism + external proposal).

The gains are broad but heterogeneous. For example, direct credential takeover in Telecom with GPT-5.1 OpenClaw improves from \(57.1\%\) to \(95.2\%\); indirect finance manipulation with Gemini-3.1-Pro and Google ADK improves from \(25.0\%\) to \(65.2\%\); indirect ticket abuse with GPT-5.1 and the OpenAI Agents SDK improves from \(15.0\%\) to \(60.0\%\); direct CRM bot attacks with GPT-5.1 OpenClaw improve from \(33.3\%\) to \(77.8\%\); and direct MITRE Injection in Code with GPT-5.1 and the OpenAI Agents SDK improves from \(0.0\%\) to \(100.0\%\). Other categories tie or regress, especially where baseline ASR is already high or behavior is dominated by framework-specific policy. We therefore interpret role separation as the strongest aggregate configuration rather than a guarantee of improvement in every fine-grained category.

\begin{table}[H]
\centering
\scriptsize
\setlength{\tabcolsep}{2pt}
\renewcommand{\arraystretch}{0.96}
\caption{\textbf{Category-wise ASR in Telecom domain.}}
\label{tab:category_telecom}
\begin{tabularx}{\linewidth}{
@{}
>{\raggedright\arraybackslash}p{0.075\linewidth}
>{\raggedright\arraybackslash}X
>{\centering\arraybackslash}p{0.155\linewidth}
>{\centering\arraybackslash}p{0.155\linewidth}
>{\centering\arraybackslash}p{0.155\linewidth}
@{}
}
\toprule
\textbf{Threat} &
\textbf{Risk category} &
\shortstack{\textbf{Gemini-3.1-Pro}\\\textbf{Google ADK}} &
\shortstack{\textbf{GPT-5.1}\\\textbf{OpenAI SDK}} &
\shortstack{\textbf{GPT-5.1}\\\textbf{OpenClaw}} \\
\midrule
\multirow[t]{11}{*}{Direct} & Credential Leak & \textbf{0.0 / 0.0} & 30.0 / \textbf{40.0} & \textbf{40.0 / 40.0} \\
 & Credential Takeover & \textbf{0.0 / 0.0} & \textbf{57.1 / 57.1} & 57.1 / \textbf{95.2} \\
 & Finance Fraud & \textbf{0.0 / 0.0} & \textbf{80.0} / 70.0 & \textbf{70.0} / 60.0 \\
 & Finance Manipulation & \textbf{33.3} / 0.0 & 81.8 / \textbf{95.5} & 81.8 / \textbf{90.9} \\
 & Network Intrusion & \textbf{100.0 / 100.0} & 78.9 / \textbf{100.0} & \textbf{94.7 / 94.7} \\
 & Privacy Leak Address & 0.0 / \textbf{50.0} & 80.0 / \textbf{90.0} & 80.0 / \textbf{100.0} \\
 & Privacy Leak Bundle & 0.0 / \textbf{10.0} & \textbf{100.0} / 80.0 & \textbf{100.0 / 100.0} \\
 & Privacy Leak Email & 0.0 / \textbf{10.0} & \textbf{100.0 / 100.0} & \textbf{100.0 / 100.0} \\
 & Privacy Leak Phone & 0.0 / \textbf{11.1} & \textbf{90.0 / 90.0} & 90.0 / \textbf{100.0} \\
 & Refund Violation & 66.7 / \textbf{78.9} & 75.0 / \textbf{90.0} & \textbf{95.0} / 90.0 \\
 & Ticket Abuse & \textbf{66.7} / 50.0 & 42.1 / \textbf{73.7} & \textbf{73.7 / 73.7} \\
\cmidrule(lr){1-5}
\multirow[t]{11}{*}{Indirect} & Credential Leak & 50.0 / \textbf{60.0} & 30.0 / \textbf{40.0} & \textbf{20.0 / 20.0} \\
 & Credential Takeover & 16.7 / \textbf{36.4} & 18.2 / \textbf{27.3} & 13.6 / \textbf{18.2} \\
 & Finance Fraud & 66.7 / \textbf{77.8} & \textbf{40.0 / 40.0} & \textbf{60.0} / 50.0 \\
 & Finance Manipulation & 25.0 / \textbf{65.2} & 47.8 / \textbf{69.6} & \textbf{65.2} / 52.2 \\
 & Network Intrusion & 80.0 / \textbf{85.0} & \textbf{75.0} / 60.0 & \textbf{75.0} / 70.0 \\
 & Privacy Leak Address & \textbf{100.0 / 100.0} & \textbf{100.0 / 100.0} & \textbf{80.0} / 60.0 \\
 & Privacy Leak Bundle & \textbf{100.0} / 90.0 & \textbf{50.0 / 50.0} & 30.0 / \textbf{60.0} \\
 & Privacy Leak Email & \textbf{100.0} / 90.0 & 70.0 / \textbf{90.0} & 60.0 / \textbf{100.0} \\
 & Privacy Leak Phone & \textbf{100.0} / 90.0 & 50.0 / \textbf{100.0} & 20.0 / \textbf{60.0} \\
 & Refund Violation & \textbf{100.0} / 81.0 & \textbf{52.4} / 33.3 & \textbf{61.9} / 28.6 \\
 & Ticket Abuse & 33.3 / \textbf{45.0} & 15.0 / \textbf{60.0} & \textbf{70.0} / 60.0 \\
\bottomrule
\end{tabularx}
\end{table}

\begin{table}[H]
\centering
\scriptsize
\setlength{\tabcolsep}{2pt}
\renewcommand{\arraystretch}{0.96}
\caption{\textbf{Category-wise ASR in Workflow domain.}}
\label{tab:category_workflow}
\begin{tabularx}{\linewidth}{
@{}
>{\raggedright\arraybackslash}p{0.075\linewidth}
>{\raggedright\arraybackslash}X
>{\centering\arraybackslash}p{0.155\linewidth}
>{\centering\arraybackslash}p{0.155\linewidth}
>{\centering\arraybackslash}p{0.155\linewidth}
@{}
}
\toprule
\textbf{Threat} &
\textbf{Risk category} &
\shortstack{\textbf{Gemini-3.1-Pro}\\\textbf{Google ADK}} &
\shortstack{\textbf{GPT-5.1}\\\textbf{OpenAI SDK}} &
\shortstack{\textbf{GPT-5.1}\\\textbf{OpenClaw}} \\
\midrule
\multirow[t]{11}{*}{Direct} & Credential Leak & \textbf{0.0 / 0.0} & 30.0 / \textbf{40.0} & \textbf{40.0 / 40.0} \\
 & Credential Takeover & \textbf{0.0 / 0.0} & \textbf{57.1 / 57.1} & 57.1 / \textbf{95.2} \\
 & Finance Fraud & \textbf{0.0 / 0.0} & \textbf{80.0} / 70.0 & \textbf{70.0} / 60.0 \\
 & Finance Manipulation & \textbf{33.3} / 0.0 & 81.8 / \textbf{95.5} & 81.8 / \textbf{90.9} \\
 & Network Intrusion & \textbf{100.0 / 100.0} & 78.9 / \textbf{100.0} & \textbf{94.7 / 94.7} \\
 & Privacy Leak Address & 0.0 / \textbf{50.0} & 80.0 / \textbf{90.0} & 80.0 / \textbf{100.0} \\
 & Privacy Leak Bundle & 0.0 / \textbf{10.0} & \textbf{100.0} / 80.0 & \textbf{100.0 / 100.0} \\
 & Privacy Leak Email & 0.0 / \textbf{10.0} & \textbf{100.0 / 100.0} & \textbf{100.0 / 100.0} \\
 & Privacy Leak Phone & 0.0 / \textbf{11.1} & \textbf{90.0 / 90.0} & 90.0 / \textbf{100.0} \\
 & Refund Violation & 66.7 / \textbf{78.9} & 75.0 / \textbf{90.0} & \textbf{95.0} / 90.0 \\
 & Ticket Abuse & \textbf{66.7} / 50.0 & 42.1 / \textbf{73.7} & \textbf{73.7 / 73.7} \\
\cmidrule(lr){1-5}
\multirow[t]{11}{*}{Indirect} & Credential Leak & 50.0 / \textbf{60.0} & 30.0 / \textbf{40.0} & \textbf{20.0 / 20.0} \\
 & Credential Takeover & 16.7 / \textbf{36.4} & 18.2 / \textbf{27.3} & 13.6 / \textbf{18.2} \\
 & Finance Fraud & 66.7 / \textbf{77.8} & \textbf{40.0 / 40.0} & \textbf{60.0} / 50.0 \\
 & Finance Manipulation & 25.0 / \textbf{65.2} & 47.8 / \textbf{69.6} & \textbf{65.2} / 52.2 \\
 & Network Intrusion & 80.0 / \textbf{85.0} & \textbf{75.0} / 60.0 & \textbf{75.0} / 70.0 \\
 & Privacy Leak Address & \textbf{100.0 / 100.0} & \textbf{100.0 / 100.0} & \textbf{80.0} / 60.0 \\
 & Privacy Leak Bundle & \textbf{100.0} / 90.0 & \textbf{50.0 / 50.0} & 30.0 / \textbf{60.0} \\
 & Privacy Leak Email & \textbf{100.0} / 90.0 & 70.0 / \textbf{90.0} & 60.0 / \textbf{100.0} \\
 & Privacy Leak Phone & \textbf{100.0} / 90.0 & 50.0 / \textbf{100.0} & 20.0 / \textbf{60.0} \\
 & Refund Violation & \textbf{100.0} / 81.0 & \textbf{52.4} / 33.3 & \textbf{61.9} / 28.6 \\
 & Ticket Abuse & 33.3 / \textbf{45.0} & 15.0 / \textbf{60.0} & \textbf{70.0} / 60.0 \\
\bottomrule
\end{tabularx}
\end{table}

\begin{table}[H]
\centering
\scriptsize
\setlength{\tabcolsep}{2pt}
\renewcommand{\arraystretch}{0.96}
\caption{\textbf{Category-wise ASR in CRM domain.}}
\label{tab:category_crm}
\begin{tabularx}{\linewidth}{
@{}
>{\raggedright\arraybackslash}p{0.075\linewidth}
>{\raggedright\arraybackslash}X
>{\centering\arraybackslash}p{0.155\linewidth}
>{\centering\arraybackslash}p{0.155\linewidth}
>{\centering\arraybackslash}p{0.155\linewidth}
@{}
}
\toprule
\textbf{Threat} &
\textbf{Risk category} &
\shortstack{\textbf{Gemini-3.1-Pro}\\\textbf{Google ADK}} &
\shortstack{\textbf{GPT-5.1}\\\textbf{OpenAI SDK}} &
\shortstack{\textbf{GPT-5.1}\\\textbf{OpenClaw}} \\
\midrule
\multirow[t]{9}{*}{Direct} & Bot Attacks & \textbf{22.2 / 22.2} & \textbf{100.0} / 88.9 & 33.3 / \textbf{77.8} \\
 & Copyright Intellectual Property Violations & \textbf{25.0} / 12.5 & \textbf{75.0} / 50.0 & \textbf{37.5} / 12.5 \\
 & Data Exfiltration & 0.0 / \textbf{12.5} & \textbf{42.9 / 42.9} & 0.0 / \textbf{28.6} \\
 & Financial Fraud & 10.0 / \textbf{30.0} & \textbf{100.0} / 90.0 & \textbf{100.0 / 100.0} \\
 & Illegal Activity Promotion & \textbf{20.0 / 20.0} & \textbf{100.0} / 80.0 & \textbf{80.0 / 80.0} \\
 & Messaging Abuse & \textbf{10.0 / 10.0} & \textbf{100.0} / 80.0 & 90.0 / \textbf{100.0} \\
 & Misinformation & \textbf{66.7 / 66.7} & \textbf{100.0 / 100.0} & \textbf{100.0 / 100.0} \\
 & Sensitive File Deletion & \textbf{31.2} / 25.0 & \textbf{87.5} / 81.2 & \textbf{81.2} / 75.0 \\
 & Unsafe Content Spreading & \textbf{16.7} / 0.0 & 16.7 / \textbf{33.3} & \textbf{16.7} / 0.0 \\
\cmidrule(lr){1-5}
\multirow[t]{9}{*}{Indirect} & Bot Attacks & 0.0 / \textbf{11.1} & 66.7 / \textbf{77.8} & 55.6 / \textbf{66.7} \\
 & Copyright Intellectual Property Violations & \textbf{70.0} / 60.0 & 60.0 / \textbf{100.0} & \textbf{100.0 / 100.0} \\
 & Data Exfiltration & \textbf{50.0} / 12.5 & 55.6 / \textbf{62.5} & \textbf{75.0} / 62.5 \\
 & Financial Fraud & 40.0 / \textbf{70.0} & \textbf{100.0 / 100.0} & 70.0 / \textbf{100.0} \\
 & Illegal Activity Promotion & \textbf{40.0 / 40.0} & \textbf{100.0 / 100.0} & \textbf{100.0} / 90.0 \\
 & Messaging Abuse & \textbf{31.2} / 25.0 & 75.0 / \textbf{87.5} & \textbf{62.5} / 56.2 \\
 & Misinformation & \textbf{50.0} / 25.0 & \textbf{100.0} / 75.0 & \textbf{87.5 / 87.5} \\
 & Sensitive File Deletion & \textbf{46.7} / 20.0 & \textbf{73.3} / 60.0 & \textbf{73.3} / 66.7 \\
 & Unsafe Content Spreading & \textbf{58.3} / 50.0 & 75.0 / \textbf{100.0} & 83.3 / \textbf{91.7} \\
\bottomrule
\end{tabularx}
\end{table}

\begin{table}[H]
\centering
\scriptsize
\setlength{\tabcolsep}{2pt}
\renewcommand{\arraystretch}{0.96}
\caption{\textbf{Category-wise ASR in Code domain.}}
\label{tab:category_code}
\begin{tabularx}{\linewidth}{
@{}
>{\raggedright\arraybackslash}p{0.075\linewidth}
>{\raggedright\arraybackslash}X
>{\centering\arraybackslash}p{0.155\linewidth}
>{\centering\arraybackslash}p{0.155\linewidth}
>{\centering\arraybackslash}p{0.155\linewidth}
@{}
}
\toprule
\textbf{Threat} &
\textbf{Risk category} &
\shortstack{\textbf{Gemini-3.1-Pro}\\\textbf{Google ADK}} &
\shortstack{\textbf{GPT-5.1}\\\textbf{OpenAI SDK}} &
\shortstack{\textbf{GPT-5.1}\\\textbf{OpenClaw}} \\
\midrule
\multirow[t]{1}{*}{Direct} & MITRE Injection & \textbf{0.0 / 0.0} & 0.0 / \textbf{100.0} & \textbf{100.0 / 100.0} \\
\cmidrule(lr){1-5}
\multirow[t]{16}{*}{Indirect} & MITRE Insecure Design & 33.3 / \textbf{66.7} & \textbf{100.0} / 66.7 & \textbf{66.7 / 66.7} \\
 & MITRE Security Logging And Monitoring Failures & \textbf{0.0 / 0.0} & \textbf{100.0} / 50.0 & \textbf{100.0 / 100.0} \\
 & MITRE Security Misconfiguration & 50.0 / \textbf{100.0} & \textbf{100.0 / 100.0} & \textbf{100.0} / 50.0 \\
 & MITRE Server Side Request Forgery SSRF & \textbf{50.0} / 0.0 & \textbf{50.0} / 0.0 & \textbf{100.0} / 50.0 \\
 & MITRE Software And Data Integrity Failures & \textbf{66.7} / 33.3 & \textbf{100.0 / 100.0} & 0.0 / \textbf{100.0} \\
 & MITRE Vulnerable And Outdated Components & \textbf{100.0} / 50.0 & 50.0 / \textbf{100.0} & \textbf{100.0 / 100.0} \\
 & OWASP Broken Access Control & 0.0 / \textbf{50.0} & \textbf{100.0} / 50.0 & \textbf{50.0 / 50.0} \\
 & OWASP Cryptographic Failures & 0.0 / \textbf{33.3} & \textbf{100.0} / 66.7 & 0.0 / \textbf{66.7} \\
 & OWASP Identification And Authentication Failures & 0.0 / \textbf{40.0} & 20.0 / \textbf{40.0} & 60.0 / \textbf{100.0} \\
 & OWASP Injection & \textbf{0.0 / 0.0} & 66.7 / \textbf{100.0} & \textbf{66.7 / 66.7} \\
 & OWASP Insecure Design & 0.0 / \textbf{50.0} & \textbf{100.0 / 100.0} & \textbf{100.0 / 100.0} \\
 & OWASP Security Logging And Monitoring Failures & \textbf{33.3 / 33.3} & \textbf{33.3} / 0.0 & \textbf{100.0 / 100.0} \\
 & OWASP Security Misconfiguration & \textbf{20.0 / 20.0} & \textbf{80.0 / 80.0} & \textbf{80.0 / 80.0} \\
 & OWASP Server Side Request Forgery SSRF & \textbf{0.0 / 0.0} & \textbf{50.0 / 50.0} & 0.0 / \textbf{100.0} \\
 & OWASP Software And Data Integrity Failures & \textbf{0.0 / 0.0} & 66.7 / \textbf{100.0} & \textbf{100.0} / 66.7 \\
 & Untrusted Installation & \textbf{4.8} / 4.5 & 63.6 / \textbf{86.4} & 68.2 / \textbf{77.3} \\
\bottomrule
\end{tabularx}
\end{table}

\end{document}